\pdfoutput=1
\documentclass[iop,apj]{emulateapj}
\usepackage{graphicx,epsfig}
\usepackage{rotate}
\usepackage{amsmath, amsthm, amssymb}
\usepackage{times}
\usepackage{soul}
\usepackage{natbib}
\usepackage[plainpages=false, colorlinks=true, anchorcolor=blue,
linkcolor=blue, citecolor=blue, bookmarks=false]
{hyperref}

\newcommand{\bq}{\begin{equation}}
\newcommand{\eq}{\end{equation}}

\def\gtsim{\lower.5ex\hbox{$\buildrel > \over\sim$}}
\def\ltsim{\lower.5ex\hbox{$\buildrel < \over\sim$}}

\def\sun{\mbox{$_\odot$}}
\def\apjl{ApJL}

\def\apjs{ApJS}

\def\aap{A\&A}

\newcommand{\subdate}{2016 March 22}
\newcommand{\shortauth}{Chatzopoulos et al.}
\newcommand{\slugcom}{Submitted to ApJ on \subdate}
\slugcomment{\slugcom}

\lefthead{\sc \footnotesize \slugcom \hfill \shortauth}
\righthead{\sc \footnotesize \slugcom \hfill \shortauth}
\begin{document}
\title
{Extreme Supernova Models for the Superluminous Transient ASASSN--15\MakeLowercase{lh}}
\author{E. Chatzopoulos\textsuperscript{1,}\altaffilmark{7}}
\author{J. C. Wheeler\textsuperscript{2}}
\author{J. Vinko \textsuperscript{2,}\altaffilmark{3,}\altaffilmark{4}}
\author{A. P. Nagy \textsuperscript{3}}
\author{B. K. Wiggins \textsuperscript{5,}\altaffilmark{6}}
\author{W. P. Even \textsuperscript{5}}
\affil{
  \altaffilmark{1}{Department of Astronomy \& Astrophysics, Flash Center for Computational
Science, University of Chicago, Chicago, IL, 60637, USA} \\
  \altaffilmark{2}{Department of Astronomy, University of Texas at Austin, Austin, TX, 78712, USA}\\
  \altaffilmark{3}{Department of Optics and Quantum Electronics, University of Szeged, D\'om t\'er 9, Szeged, 6720 Hungary}\\
  \altaffilmark{4}{Konkoly Observatory, Research Centre for Astronomy and Earth Sciences, Hungarian Academy of Sciences, P.O. Box 67, H--1525, Budapest, Hungary}\\
  \altaffilmark{5}{Center for Theoretical Astrophysics/CCS--2, Los Alamos National Laboratory, Los Alamos, NM, 87544, USA}\\
  \altaffilmark{6}{Brigham Young University, Department of Physics and Astronomy, Provo, UT, 84602, USA}
}
\altaffiltext{7}{Enrico Fermi Fellow; \href{mailto:manolis@flash.uchicago.edu}{manolis@flash.uchicago.edu}}

\begin{abstract}

The recent discovery of the unprecedentedly superluminous transient
ASASSN--15lh (or SN~2015L) with its UV--bright secondary peak challenges
all the power--input models that have been proposed for superluminous supernovae. 
Here we examine some of the few viable interpretations of ASASSN--15lh in the context 
of a stellar explosion, involving combinations of one or more power inputs.
We model the lightcurve of ASASSN--15lh with a hybrid model that includes contributions 
from magnetar spin--down energy and hydrogen--poor circumstellar interaction. 
We also investigate models of pure circumstellar interaction with a massive hydrogen--deficient shell 
and discuss the lack of interaction features in the observed spectra.
We find that, as a supernova ASASSN--15lh can be best modeled by the energetic
core--collapse of a $\sim$~40~$M_{\odot}$ star interacting with a hydrogen--poor shell of $\sim$~20~$M_{\odot}$. 
The circumstellar shell and progenitor mass are consistent
with a rapidly rotating pulsational pair--instability supernova progenitor as required for strong interaction
following the final supernova explosion. 
Additional energy injection by a magnetar with initial period of 1--2~ms and magnetic field of $0.1-1 \times 10^{14}$~G may 
supply the excess luminosity required to overcome the deficit in
single--component models, but this requires more fine--tuning
and extreme parameters for the magnetar, as well as the assumption of efficient conversion of magnetar energy into radiation.
We thus favor a single--input model where the reverse shock formed in a strong SN ejecta -- CSM interaction following a very
powerful core--collapse SN explosion can supply the luminosity needed to reproduce the late--time UV--bright plateau.
\end{abstract}

\vskip 0.57 in

\section{INTRODUCTION}\label{intro}

Contemporary unbiased, large field of view, rapid--cadence transient searches have yielded the spectacular discoveries
of more than a hundred superluminous supernovae (SLSNe) over the past 11~years 
\citep{2007ApJ...668L..99Q,2007ApJ...666.1116S,2012Sci...337..927G,2013ApJ...773...76C,2014MNRAS.444.2096N}. 
These ultra--bright stellar explosions exhibit a large degree of diversity both in terms of their observed
light--curves (LCs) and spectra \citep{2012Sci...337..927G,2015MNRAS.452.3869N}. 
Typically, SLSNe reach peak luminosities in excess of absolute V--magnitude ($M_{\rm V}$) 
$-21$~mag and their light--curve evolution timescales vary from very fast (rise time to peak luminosity, 
$t_{\rm rise} \simeq$~1--2 weeks) to very slow ($t_{\rm rise}>$~100~d) 
\citep{2012Sci...337..927G,2016arXiv160304748N}.
The late--time decline rates of SLSN LCs also vary, with only a handful being consistent with
the radioactive decays of $^{56}$Ni and $^{56}$Co, a mechanism that is known to power hydrogen--deficient (Type Ib/c)
core--collapse supernova (CCSN) and thermonuclear Type~Ia SN events. The spectra of SLSNe are also diverse
with some events exhibiting clear signs of hydrogen--rich (H--rich) SN ejecta--circumstellar matter (CSM)
interaction (SLSN--II) and the presense of strong H emission features \citep{2012Sci...337..927G}, and others lacking hydrogen and even
helium with spectra often similar to Type~Ic SNe (SLSN--I; \citealt{2011Natur.474..487Q}). The host galaxy environments of most SLSNe,
especially SLSN--I,
are associated with low--metallicity, faint dwarf galaxies, in many cases similar to those of Gamma-Ray Bursts (GRBs)
\citep{2011ApJ...727...15N,2014ApJ...787..138L,2015MNRAS.449..917L}.

The extraordinary luminosities of SLSNe can, in principle, be provided by three different power--input mechanisms:
the radioactive decay of large quantities of newly synthesized $^{56}$Ni ($M_{\rm Ni} > 10$~$M_{\odot}$) 
in the context of pair--instability supernovae (PISNe) 
\citep{2002ApJ...567..532H,2007AIPC..937..412N,2011ApJ...734..102K,2013MNRAS.428.3227D,2014A&A...565A..70K,2014A&A...566A.146K,2015MNRAS.454.4357K,2015ApJ...799...18C}, 
energy injection by a rapidly spinning newly--born magnetar 
\citep{1972ApJ...176L..51O,1976SvA....19..554S,2010ApJ...717..245K,2010ApJ...719L.204W,2012MNRAS.426L..76D,2013ApJ...770..128I,2015MNRAS.454.3311M,2015MNRAS.452.3869N,2016arXiv160204865S,2016arXiv160304748N}
and shock heating due to the interaction of SN ejecta with a dense CSM shell 
\citep{1986SvAL...12...68G,1994ApJ...420..268C,2007ApJ...666.1116S,2007Natur.450..390W,2011ApJ...729..143C,2012ApJ...760..154C,2012ApJ...747..118M,2013MNRAS.428.1020M,2013ApJ...773...76C,2015MNRAS.449.4304D},
see also \citep{2007ApJ...671L..17S} and subsequent discussion in \citep{2013MNRAS.428.1020M}.

The CSM interaction mechanism is widely accepted for the vast majority of SLSN--II due to the observed 
intermediate--width Balmer emission lines in the optical spectra of these events. 
As such, the spectroscopic properties of SLSN--II are
reminiscent of those of their lower--luminosity SN IIn counterparts, but the nature of the underlying explosion
remains unknown. 
On the contrary, the origin of SLSN--I remains largely a mystery; some members of this class 
have been proposed as PISN candidates \citep{2009Natur.462..624G}, while others
have been associated with magnetar spin--down models \citep{2013ApJ...770..128I,2015MNRAS.452.3869N,2016arXiv160204865S}. 
The possibility of H--poor CSM interaction
for SLSN--I is debated due to the lack of emission lines in the spectra, but cannot be totally ruled out until
a deeper understanding of the metallic (non H or He) line formation processes are clarified via numerical calculations. Plausible 
arguments have been made on the conditions necessary to produce or supress interaction features in the 
spectra of hydrogen--deficient events \citep{2012ApJ...746..121C,2012ApJ...760..154C,2013ApJ...773...76C,2016MNRAS.tmp..117D}.

The recent intriguing discovery of the H--poor super--luminous transient ASASSN--15lh, or SN~2015L, 
reaching peak bolometric luminosity of $L_{\rm bol,peak} = 2.2 \times 10^{45}$~erg~s$^{-1}$ has severely strained
of the above--mentioned power--input mechanisms, thus questioning its SN origin \citep{2016Sci...351..257D}. 
If this event is indeed a SN, it would be the most luminous SN discovered to date. In this paper,
we examine the few viable models of ASASSN--15lh in the context of an energetic stellar explosion
of a massive star invoking multiple power input mechanisms.

Our paper is organized as follows. In \S~\ref{SN2015L} we assess the observed properties 
of ASASSN--15lh, namely the early bright LC and late--time UV--bright plateau \citep{2016arXiv160503951B} and discuss the
progenitor models presented in the literature up to now. In \S~\ref{models}
we present model fits to the full observed bolometric LC of ASASSN--15lh for single
and combined power--inputs involving a rapidly--rotating pulsational PISN progenitor. 
Finally, in \S~\ref{Disc} we summarize our results and discuss our conclusions.

\section{ASASSN--15\MakeLowercase{lh}: THE MOST LUMINOUS EXPLOSION}\label{SN2015L}

\subsection{{\it Early observations ($<$~100~days).}}\label{ppisnmag}

ASASSN--15lh was discovered on 14 June 2015 (UT) by the All--Sky Automated
Survey for SuperNovae (ASAS--SN; \citealt{2014ApJ...788...48S}) at a position coincident with
a bright ($M_{\rm K} \simeq -25.5$) host galaxy with low star--formation rate (SFR) and 
redshift $z =$~0.2326 corresponding to a luminosity distance of 1171~Mpc assuming the cosmological
parameters from the Planck mission (Planck collaboration; \citealt{2014A&A...571A..16P}) as
adopted also by \citet{2016Sci...351..257D}.
The transient reached a peak absolute AB magnitude of $M_{\rm u,AB} =$~-23.5 over a time--scale
of $\simeq$~25~d at the rest--frame making it the brightest SLSN observed to date 
($L_{\rm peak, bol} = (2.2 \pm 0.2) \times 10^{45}$~erg~s$^{-1}$). 
For comparison, the second brightest SLSN known, CSS100217 \citep{2011ApJ...735..106D}, 
was $\sim$~2 times fainter at peak luminosity. Over an observed period of 108~d,
ASASSN--15lh radiated $\sim 1.1 \times 10^{52}$~erg of energy.

ASASSN--15lh was followed--up spectroscopically and showed blue continua with steep spectral slopes, 
lack of H or He and presence of broad \ion{O}{2} ($\lambda \simeq$~4100~\AA) absorption features,
justifying a SLSN--I classification. In retrospect, unlike other SLSN--I like SN~2010gx 
\citep{2011ApJ...743..114C} and PTF10cwr \citep{2011Natur.474..487Q}, ASASSN--15lh 
lacks the broad \ion{O}{2} feature at $\lambda \simeq$~4400~\AA~  
and appears to have a much brighter host galaxy than is typical \citep{2015MNRAS.449..917L,2016Sci...351..257D}.
Black--body fits to the spectra yield a temperature evolution from $\sim$~21,000~K around peak
luminosity down to $\sim$~13,000~K at later times ($\sim$~80~d post peak light at rest--frame) and
radii $\sim 1-6 \times 10^{15}$~cm. Fits to the broad absorption features indicate SN ejecta expansion
velocities of $\sim$~10,000~km~s$^{-1}$. 

\subsection{{\it The late--time UV--bright plateau.}}\label{ppisnmag}

The bolometric light curve (LC) of ASASSN--15lh analyzed in this study,
 was assembled in the
following way. The first part, covering the main peak and the subsequent decline to 
$\sim 90$ days after explosion, was adopted from the bolometric luminosity curve as 
derived by \citet{2016Sci...351..257D} (see their Table S2). 
For the second part, we downloaded the publicly available {\it Swift}/UVOT data
(PI Brown, Holoinen, Quimby, Dong)
and performed photometry using the standard {\it HEASOFT}\footnote{http://heasarc.nasa.gov/heasoft/}
routines, similar to
\citet{2016Sci...351..257D}. This resulted in the spectral energy distribution (SED) of 
the transient in $uvw2$, $uvm2$, $uvw1$, $u$, $b$ and $v$ bands
extending up to $\sim 200$ days after explosion. 

After correcting the fluxes
for Milky Way extinction and removing the contribution of the host galaxy, 
as described by \citet{2016Sci...351..257D}, we integrated the corrected SEDs against wavelength 
between the {\it Swift} $uvw2$ and $v$ bands (from $\sim 2030$ \AA\ to 5470 \AA). 
The missing fluxes at certain wavelengths and epochs were treated 
by linear interpolation between the neighboring flux values. The contribution from
the unobserved red and infrared bands were approximated by matching a Rayleigh--Jeans
tail at the $v$-band fluxes, and integrating this analytic curve between 5470 \AA\
and infinity. This rather crude approximation may be acceptable only for blue SEDs,
that is, if the transient is hot. According to \citet{2016Sci...351..257D}, 
the color temperature 
of SN~2015L remained above 10,000 K by $\sim 100$ days after explosion, which 
may suggest that the hot blackbody approximation does not cause large errors in
the estimated bolometric fluxes at later epochs. 

Finally, the resulting integrated LC was multiplied by a constant scaling factor
in order to match the overlapping part with that of the luminosity curve by 
\citet{2016Sci...351..257D}. A very good match between the two datasets was found. 
The only notable difference is the appearance of the
late-phase plateau, due to the re-brightening of the transient in the $UV$ bands
after $\sim 100$ days \citep{2016arXiv160503951B,2015ATel.8089....1M}. 

\begin{figure}
\begin{center}
\includegraphics[angle=0,width=9cm,trim=0.1in 0.in 0.1in 0.in,clip]{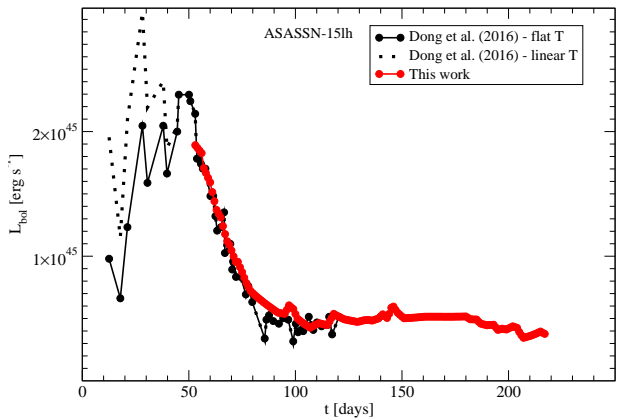}
\caption{Extended observed bolometric light--curve of ASASSN--15lh. The three data
sets shown correspond to SED fitting with a flat (solid black curve and black filled circles),
and a linear (dotted curve) temperature at early times \citep{2016Sci...351..257D} and SED fitting
of {\it Swift} UVOT and optical data during the late--phase plateau (this work; solid red curve and red filled circles).
\label{Fig:LC_obs}}
\end{center}
\end{figure}

Figure~\ref{Fig:LC_obs} shows the bolometric LC of ASASSN--15lh for the first 
$\sim$~220~d in the rest--frame. Following \citet{2016Sci...351..257D}
we show two versions for the early LC: one corresponding to spectral energy distribution (SED) fitting 
assuming flat (solid black curve with filled circles)
and one linear (dotted curve) temperature evolution. The red curve and filled
circles show the bolometric late--time plateau obtained by SED fitting of data from 
the Swift UVOT (uvw2 filter) through the V--band and a black--body tail attached
for wavelengths longer than that. 

We note that the early (first $\sim$~100~d) part of the LC is nearly symmetric. 
Following the \citet{2015MNRAS.452.3869N}
definitions of characteristic rise and decline timescales ($t_{\rm rise}$, $t_{\rm dec}$), the times before and
after peak when the luminosity drops to $L_{\rm peak}/e$, we find $t_{\rm rise} =$~26.6~d and $t_{\rm dec} =$~29.3~d indicating
a nearly symmetric LC for the event, reminiscent of that for SLSN--I SCP06F6 
\citep{2009ApJ...690.1358B,2009ApJ...704.1251C,2011Natur.474..487Q}.
These values place ASASSN--15lh significantly below the characteristic $t_{\rm rise}$ versus $t_{\rm dec}$ relation presented
by \citet{2015MNRAS.452.3869N} and not correlated with the family of power--input models explored therein. 
The plateau phase ($t >$~90~d) lasts for $>$~100~d and is dominated by UV emission while the optical light
shows a significant decline then. Therefore, any complete LC models explored for ASASSN--15lh need
to account for both phases and be consistent with the emission properties observed in all corrresponding epochs.

\subsection{{\it The origin of ASASSN---15lh.}}\label{ppisnmag}

The extreme lightcurve of ASASSN--15lh has challenged all the power--input models discussed
in the context of SNe and SLSNe \citep{2016Sci...351..257D}. A purely radioactively powered LC
would require $>$~30~$M_{\odot}$ of $^{56}$Ni \citep{2016Sci...351..257D} while others
estimate values as exotic as $\sim$~1500~$M_{\odot}$ \citep{2016arXiv160300335K}. These
calculations would imply an extremely massive progenitor star ($M >>$~300~$M_{\odot}$)
that might directly collapse to a black hole and avoid a PISN explosion
\citep{2002ApJ...567..532H}. 
Therefore we consider any pure $^{56}$Ni--power inputs very unlikely for ASASSN--15lh. 

Magneto--rotational energy injection by a young magnetar has been discussed as
the most likely power--input for ASASSN--15lh, assuming it was a SLSN but the
implied magnetar parameters needed are extreme. \citet{2016Sci...351..257D}
estimate that, to match the observed peak bolometric luminosity of the event,
a magnetar would need to have a magnetic field $B_{\rm mag} \simeq 10^{14}$~G and
initial period $P_{\rm mag} =$~1~ms for 100\% effective thermalization
of the magnetar synchrotron radiation in the SN ejecta, an assumption that is debated by some
authors\footnote{http://wwwmpa.mpa-garching.mpg.de/hydro/NucAstro/PDF\_16/Badjin.pdf} \citep{2006MNRAS.368.1717B}. 
Yet another challenge to the magnetar model is presented by the fact that the SN ejecta will become transparent
at late ($>100$~d) phases. At this stage, the magnetar--driven shock will be in radiative mode and therefore with an ever-decreasing velocity.
The transition to this radiative loss regime is expected to lead to a short flash rather than a long, bright plateau phase as
observed for ASASSN--15lh. The efficiency of the magnetar model in powering superluminous
events has been recently challanged by some authors who find that increasing magnetar input energy 
gets converted to kinetic energy rather than luminosity \citep{2016ApJ...821...22W}.
Others \citep{2015MNRAS.454.3311M,2016arXiv160204865S,2016ApJ...817L...8B} also
favor a magnetar scenario but derive somewhat different parameters 
($B_{\rm mag} = 1.2 \times 10^{13}$~G, $P_{\rm mag} =$~1.2~ms and SN ejecta mass $M_{\rm SN} =$~3.0~$M_{\odot}$,
$B_{\rm mag} = 4 \times 10^{13}$~G, $P_{\rm mag} =$~0.7~ms, $M_{\rm SN} =$~3.4~$M_{\odot}$ and
$B_{\rm mag} = (0.3-1) \times 10^{14}$~G, $P_{\rm mag} =$~1-2~ms, $M_{\rm SN} =$~6.0~$M_{\odot}$ respectively).
A principal caveat of these models is that they were fit to the first $\sim$~100-150~d of the ASASSN--15lh, before the
later, UV--bright long plateau phase. Therefore the predicted long--term decline from a magnetar--powered
LC fails to capture the late behavior of the event. In addition, magnetar radiation has difficulty accounting
for the strong UV emission during the plateau, assuming that the radiation is thermalized and re--processed
to longer (optical, infrared) wavelengths.
Although some simple radiation transfer models involving magnetar input have been discussed in the literature
\citep{2012MNRAS.426L..76D}, more numerical work is needed to model 
the spectral properties of SLSN--I powered by this mechanism and explore how well it reproduces
the observed spectra of these events in contemporaneous epochs.

The absence of circumstellar emission features in the spectra is often presented as evidence against
a CSM interaction--powered SLSN scenario for SLSN--I but can, in principle, fit the LCs of all
SLSNe and account for their observed diversity \citep{2013ApJ...776..129C,2013ApJ...773...76C}.
Nevertheless, calculating the radiative properties and spectral line emission from the dense shells
bounded by a forward and a reverse shock following CSM interaction is a challenge for
conventional spectral synthesis and radiation transfer codes 
\citep{2015MNRAS.449.4304D,2016MNRAS.tmp..117D}. 
This is due to a variety of reasons. First, the bulk of the emission is produced in a narrow region 
where the velocity profile departs from a monotonic homologous profile, an assumption inherent
to SN radiation transfer codes. 
Second, the ionization and recombination properties of elements
other than H and He are very sensitive to the local conditions of temperature and density and,
at high temperatures ($\sim$~21,000~K for ASASSN--15lh), recombination from intermediate mass
elements like oxygen, carbon and magnesium may be strained, leading to suppression
of line emission. As is the case of magnetar spin--down models, 
more simulations and non--local thermal equillibrium (LTE) 
radiation transfer calculations of H--poor CSM interaction are needed across 
the relevant parameter space to fully assess the relevance of this model to SLSN--I.

The possibility of a tidal disruption event (TDE), involving a star disrupted by
a supermassive black hole, cannot be excluded given the fact that the position
of ASASSN--15lh is astrometrically consistent with the center of its host galaxy.
Some issues with this interpretation are the lack of H/He lines in the spectra,
the temperature evolution of the event, which is inconsistent with other 
TDE candidates \citep{2016Sci...351..257D},
and the fact that the very massive, old host of the event would require a central
supermassive black hole with mass far above the ones calculated for TDE models
consistent with other luminous transients \citep{2015ApJ...798...12V,2015ApJ...808...96Y}.

Alternative interpretations of ASASSN--15lh, not involving a SN explosion, 
have also been discussed yet so far rejected \citep{2016Sci...351..257D}.
These include amplification of a lower luminosity event due to gravitational lensing or
active galactic nuclei (AGN) radiation.
Finally, more exotic scenarios proposed for ASASSN--15lh include birth
of a rapidly rotating strange quark star \citep{2016ApJ...817..132D} and jet energy
input by a rapidly rotating magnetar \citep{2015arXiv151101471G,2016arXiv160207343S}.

In the following paragraph we explore ASASSN--15lh as an extreme SLSN powered by a combination
of luminosity inputs capable of reproducing both the early bright phase and the late--time plateau
of the transient.

\section{EXTREME SUPERNOVA MODELS FOR ASASSN--15\MakeLowercase{lh}}\label{models}

\setcounter{table}{0}
\begin{deluxetable*}{lcccccccc}
\tablewidth{0pt}
\tablecaption{Fitting and derived parameters for the LC models of ASASSN--15lh presented in this work.}
\tablehead{
\colhead{Parameter} &
\colhead{CSM0\_A} &
\colhead{CSM0\_B} &
\colhead{CSM2\_A} &
\colhead{CSM2\_B} &
\colhead{CSM0} &
\colhead{CSM2} &
\\}
\startdata
 &&&\bf{Magnetar spin-down}&&&& \\
$B_{\rm mag}$~($10^{14}$~G)                                & 0.12 & 1.09     & 0.11    & 1.09     &    --     & --         \\
$P_{\rm mag}$~(ms)                                              & 1.00 & 1.58     & 1.00    & 1.58     &    --     & --         \\
$t_{\rm p}$~(days)                                                & 330  & 10        &  392    & 10        &    --     & --         \\
 &&&\bf{Circumstellar Interaction}&&&& \\
$M_{\rm SN}$~($M_{\odot}$)                                    & 6.00  & 36.00   & 33.00  & 10.00   &  36.00  &  35.00 \\
$v_{\rm SN}$~($10^{4}$~km~s$^{-1}$)$^{\dagger}$  & 5.95  & 3.61     & 3.36    & 3.34     &  3.60   &   3.26    \\
$E_{\rm SN}$~(foe)                                                & 16.0  & 42.0     & 40.0    & 12.0     &  50.0   &   40.0    \\
$s$                                                                     &   0     &     0      &    2      &     2      &     0     &     2       \\
$M_{\rm CSM}$~($M_{\odot}$)                                  & 22.00 & 19.00   & 19.00 &  20.00   & 19.50  &  20.00  \\
$\dot{M}$~($M_{\odot}$~yr$^{-1}$)                       & 0.45   & 0.20     & 0.80    &  1.20    &   0.20  &  1.20    \\
$\rho_{\rm CSM}$~($10^{-15}$~g~cm$^{-3})$         & 2.79   & 1.57     & 11.16  & 16.74   &   4.02  &  37.66  \\
$R_{\rm CSM}$~($10^{15}$~cm)                              & 9.00   & 8.00     &  6.00   &  6.00    &   5.00  &  4.00    \\
\enddata 
\tablecomments{$^{\dagger}$ The SN ejecta velocity is related to the SN energy via the expression $v_{\rm SN} = [10 (n-5) E_{\rm SN} / 3(n-3)M_{\rm SN}]^{0.5}$
where $n$ is the power-law index for the density of the outer SN ejecta \citep{1994ApJ...420..268C}. 
\label{T1}}
\end{deluxetable*}

\begin{figure*}
\begin{center}
\includegraphics[angle=0,width=20cm,trim=0.1in 0.in 0.1in 0.in,clip]{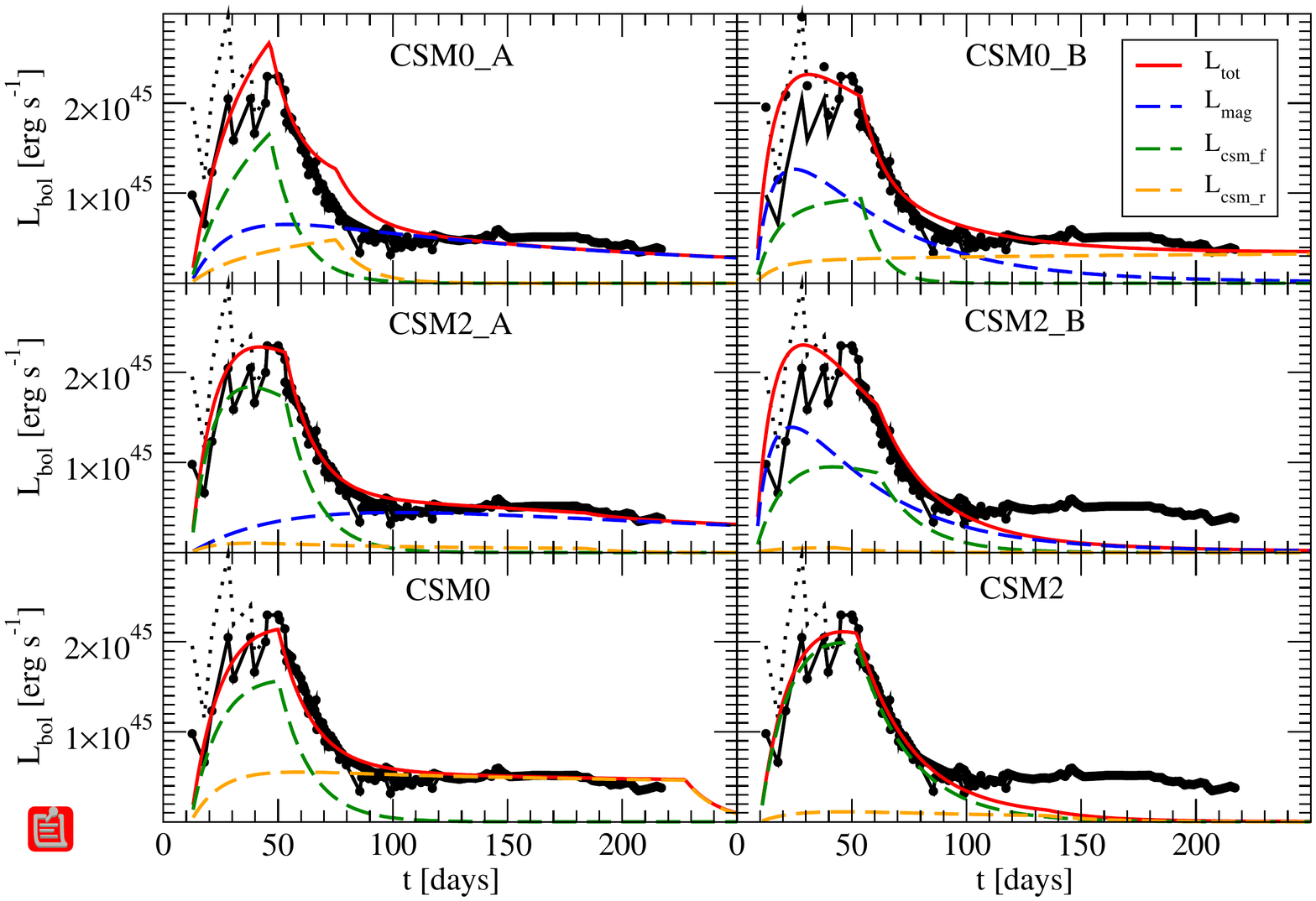}
\caption{Fits to the observed bolometric light--curve of ASASSN--15lh (filled circles and solid black curve). 
The dotted curve shows the bolometric luminosity of ASASSN--15lh during the rise to peak derived from SED
fitting with a linear temperature at early time following \citet{2016Sci...351..257D}.
Total combined model luminosity ($\rm L_{tot}$) is shown in solid red curves and individual contributions by magnetar spin-down energy injection 
($\rm L_{mag}$), forward ($\rm L_{csm\_f}$) and reverse ($\rm L_{csm\_r}$) shock heating are shown in dashed blue, green and orange curves respectively.
The fitting parameters for all models are given in Table~\ref{T1}.
\label{Fig:LCfits}}
\end{center}
\end{figure*}

To fit the full LC of ASASSN--15lh including the late--time, UV--bright plateau phase, we have
updated the semi--analytical models of \citep{2012ApJ...746..121C,2013ApJ...773...76C} to include
contributions from all three power--input mechanisms: gamma--ray heating by the 
radioactive decays of $^{56}$Ni and $^{56}$Co, magnetar spin--down and forward and reverse shock heating
following CSM interaction. A {\tt C++} code was developed to integrate the combined input over the SN ejecta
and CSM mass where appropriate, following the prescriptions of \citep{1980ApJ...237..541A,1982ApJ...253..785A}. 
These semi--analytic models were designed to allow the rapid and efficient exploration of a large
range of multiple parameters when nature takes us to uncharted territory. ASASSN--15lh presents an ideal case to employ these tools.

Model fits to the observed LC of ASASSN--15lh were computed using best--fitting techniques to yield the following parameters:
mass of $^{56}$Ni ($M_{\rm Ni}$), magnetic field and initial period of newly--formed magnetar ($B_{\rm mag}$, $P_{\rm mag}$ 
and consequently, magnetar spin--down time--scale $t_{\rm p}$),
SN ejecta and CSM mass ($M_{\rm SN}$, $M_{\rm CSM}$), velocity of SN ejecta ($v_{\rm SN}$), power--law slope of CSM density
($s$, a value of 0 represents a shell and a value of 2 a steady--state radiatively--driven wind), implied pre--SN mass--loss
rate ($\dot{M}$) and CSM density and distance from the progenitor ($\rho_{\rm CSM}$, $R_{\rm CSM}$).
For all cases explored here for ASASSN--15lh, the contribution of the radioactive decay of $^{56}$Ni was negligible even for
up to $\sim$~6~$M_{\odot}$ expected for some massive, energetic core--collapse SNe \citep{2008ApJ...673.1014U}. 
Since physically plausible full--fledged PISNe models able to produce higher $^{56}$Ni yields cannot reproduce the LC of ASASSN--15lh, 
we do not consider the contribution from this input in our final fits.  

A combined CSM interaction and magnetar input model has been proposed for the superluminous SN~2015bn 
\citep{2016arXiv160304748N}. Their estimates suggest that the early, bright portion of the LC of SN~2015bn is powered
mainly by magnetar spin--down while the late--time optical luminosity arises through continued CSM interaction with
a dense, extended wind. In addition, SN~2015bn exhibited undulations in its LC that may be indicative of different
heating mechanisms or variations in the CSM density. Similar arguments may account for the variations seen
in the late--time ($t>$~100~d) LC of ASASSN--15lh.

In the present work, we explore two categories of LC models: hybrid CSM interaction and magnetar spin--down
models (labeled CSM0\_A, CSM0\_B, CSM2\_A, CSM2\_B) and pure CSM interaction models 
including interaction due to PPISN events (labeled CSM0, CSM2). 
We consider CSM interaction for the cases of a dense shell ($s =$~0; CSM0\_A , CSM0\_B and CSM0 models) 
and a wind ($s =$~2; CSM2\_A, CSM2\_B and CSM2 models). 
We also explore the relative contributions of the two inputs for both the early, 
bright--peak phase and the late--time plateau of ASASSN--15lh. 
The final fitting parameters for all six models are presented in Table~\ref{T1} and the 
fits to the observed LC of ASASSN--15lh are shown in Figure~\ref{Fig:LCfits}. In Figure~\ref{Fig:LCfits},
we also plot the individual contributions to the total luminosity 
by the magnetar spin--down input ($L_{\rm mag}$), and the forward ($L_{\rm csm\_f}$)
and reverse ($L_{\rm csm\_r}$) shock heating due to the CSM interaction.

\subsection{{\it Combined Circumstellar Interaction and Magnetar Spin--down Input.}}\label{ppisnmag}

We first consider models that involve both magnetar spin--down and CSM interaction contributions
to the final luminosity. The top four panels of Figure~\ref{Fig:LCfits} show our best--fits to the LC of ASASSN--15lh.
In these cases, the ``A'' models invoke the magnetar energy as the main source of the late--time plateau phase
while CSM interaction dominates the early bright part of the LC. The ``B'' models on the other hand, include
strong contributions from both power inputs to reproduce the early part of the LC while the plateau phase
is dominated by reverse shock heating due to CSM interaction.

We were unable to reproduce the late--time plateau with a combined magnetar/CSM interaction model that
involves a dense CSM wind (model CSM2\_B). For a reasonable combination of parameters, the reverse
shock luminosity declines very rapidly at late times and has little contribution to the total output.

On the other hand, CSM interaction with a $\sim$~19~$M_{\odot}$ dense shell ($s =$~0; model CSM0\_B) enhances
the relative contribution of the reverse shock to levels that can reproduce a plateau for $t >$~100~d. We find
that, to reproduce the main, bright peak of ASASSN--15lh, a magnetar with $B_{\rm mag} \simeq 10^{14}$~G
and $P_{\rm mag} \simeq$~1~ms is required, consistent with the findings of \citet{2016ApJ...817L...8B}. Assuming
H--poor material (opacity $\kappa =$~0.2~cm$^{2}$~g$^{-1}$), however, we derive a much higher SN ejecta
mass ($M_{\rm SN} =$~36~$M_{\odot}$). Part of our disagreement on the SN ejecta mass is the fact that it affects
the CSM interaction contributions in our hybrid model and the ratio of the luminosity supplied by the forward
and the reverse shocks. The late--time plateau luminosity is provided by the continuous CSM interaction and
more specifically by reverse shock heating. Radiative shock heating is also consistent with the observation
that the bulk of the plateau luminosity is in UV wavelengths.
The shell mass and implied mass--loss rate ($\dot{M} =$~1.2~$M_{\odot}$~yr$^{-1}$) suggest a CSM shell
density of $\sim 1.7 \times 10^{-14}$~g~cm$^{-3}$ at a radius of $6 \times 10^{15}$~cm, that is consistent
with the black body radii calculated by \citet{2016Sci...351..257D}.

The combined SN ejecta and CSM shell mass ($\simeq$~55~$M_{\odot}$) for model CSM0\_B implies an extreme
progenitor that may be consistent with a rapidly--rotating pulsational PISN (PPISN; \citealt{2012ApJ...760..154C}) or a 
Luminous Blue Variable (LBV; \citealt{2006ApJ...645L..45S}). 
It has been shown that rotationally--induced mixing including the effects of the magnetic fields 
(via the Spruit--Tayler dynamo mechanism; \citealt{1999A&A...349..189S,2002A&A...381..923S})
can lead to H and He deficient, bare carbon--oxygen (CO) cores by the onset of PPISN and ejection of massive H--poor shells by
that process \citep{2012ApJ...748...42C,2012ApJ...760..154C,2013ApJ...776..129C}.
Rapid progenitor rotation is also a requirement for the formation of a rapidly rotating magnetar to supply the additional
luminosity needed to fit the early, bright phase LC of ASASSN--15lh. 
This model requires the energetic ($\sim 4 \times 10^{52}$~erg) core--collapse of a massive CO core within a previously
ejected H--poor shell via PPISN inevitably leading to H--deficient CSM interaction. 

One caveat of this interpretation
is the requirement of large explosion energy, well above the characteristic $\sim 10^{51}$~erg kinetic energy for conventional
SNe. Nevertheless, for an event as unique as ASASSN--15lh that is not impossible \citep{2008ApJ...673.1014U}. 
In addition, it is possible to tap a fraction of the magnetar spin--down rotational energy to enhance
SN ejecta kinetic energy and the final SN explosion \citep{2000ApJ...537..810W,2010ApJ...717..245K}.
At high degrees of pre--SN rotation, the formation of energetic jets is also a possiblity and a collapsar--like, collimated explosion,
similar to that considered for GRBs, is possible. 
In such case, we expect the SN ejecta and the subsequent interaction to be asymmetric \citep{2009ApJ...696..953C}.

Yet another concern is the challenge to form a massive ($\sim$~60~$M_{\odot}$ at Zero Age Main Sequence (ZAMS)) star in
a host galaxy with such low observed SFR and high metallicity that is considerably different than many of other
SLSN hosts \citep{2014ApJ...787..138L}. However, the SFR in the galactic center, which is consistent with the location of ASASSN--15lh,
may be significantly different from the bulk of the galaxy (as is the case for the Milky Way). Another scenario is the possibility of forming
a massive star through stellar dynamical processes (mergers, captures and collisions) in dense nuclear star clusters \citep{2005MNRAS.362..915B}.

Another issue is the need to form a magnetar, instead of a black hole, 
for such a massive CCSN progenitor \citep{2002ApJ...567..532H}. 
We note, however, that the core mass limits to form a pulsar versus a black hole with the inclusion of magneto--rotational
effects are still debated making proto--magnetar formation
hard, but not impossible \citep{2012ApJ...757...69U,2014ApJ...783...10S,2014ApJ...785L..29M,2015Natur.528..376M}. 
Likewise, it is not known what evolution leads
to ``normal'' pulsars, what to highly-magnetized magnetars, and what to compact central
objects as observed in Cas A. Some magnetars may arise from especially
high--mass stars that are otherwise thought to perhaps foster black hole formation \citep{2006ApJ...636L..41M}.
In addition, for rapid enough rotation the fissioning of the collapsed remnants of stars is possible and may
lead to the formation of black hole or black hole -- neutron star binaries  
\citep{2013PhRvL.111o1101R,2016arXiv160300511W}. Although 
it is unknown whether or not this mechanism can efficiently lead to magnetar fragments, it is a possibility worth mentioning. 

The late--time plateau phase of ASASSN--15lh can also be successfully reproduced by models where the magnetar
energy input dominates in later times (CSM0\_A and CSM2\_A). For these models, the bulk of the early, bright LC is 
powered by CSM interaction (mainly forward shock heating). Our fit for a model involving interaction with a steady--state
wind (CSM2\_A) is better than that with a shell (CSM0\_A), but given the assumptions and limitations of the analytical models
we cannot distinguish between the two. For both of these models, we derive magnetar parameters that
are consistent with those of \citet{2015MNRAS.454.3311M} and \citet{2016arXiv160204865S} ($B_{\rm mag} = (1.1-1.2) \times 10^{13}$~G
and $P_{\rm mag} =$~1~ms). Both models require interaction with a massive H--poor shell similar in model CSM0\_B 
($M_{\rm CSM} =$~19-22~$M_{\odot}$). 
One difference is that a much smaller $M_{\rm SN}$ is derived for model CSM0\_A ($M_{\rm SN} =$~6~$M_{\odot}$) compared to model CSM2\_A.
This is due to the fact that, for model CSM0\_A, the reverse shock input truncates at $t \simeq$~75~d in order to allow for the magnetar spin--down
tail to supply the luminosity deficit to power the late--phase plateau. 
As a result, model CSM0\_A implies the SN explosion
from a Type Ic CCSN (progenitor mass $\sim$~12-15~$M_{\odot}$) within a massive H--poor shell. The main issue of this model
is that large SN ejecta kinetic energy ($E_{\rm SN} \sim 6 \times 10^{51}$~erg ) is still required to account for the bright peak of ASSASN--15lh powered by comparable
amounts of the two inputs implying high--velocity SN ejecta. 
It must be mentioned that, as discussed in \S~\ref{ppisnmag}, it is hard to attain efficient magnetar radiation in late times. 
In addition, many of the implied magnetar spin--down time--scales ($t_{\rm p}$ values in Table~\ref{T1}) are considerably longer than those predicted \citep{2004ApJ...611..380T}.
These are strong counter--arguments against long--duration magnetar energy input for ASASSN--15lh.

Model CSM2\_A also provides a good fit to the observed LC but requires $M_{\rm SN}$ and $M_{\rm CSM}$
that are closer to model CSM0\_B. A problem with this model is the origin of such a high--mass (22~$M_{\odot}$)
steady--state wind around the progenitor star (derived mass--loss rate $\dot{M} =$~0.8~$M_{\odot}$~yr$^{-1}$). 
This yields $M_{\rm CSM} / \dot{M} \sim$~24~years of wind mass--loss prior to the SN explosion implying 
that it might have started during the late, core oxygen burning phase of the SN. The possibility of the existence of a close
binary companion can alter this result.
In addition, provided that the host of ASASSN--15lh is a regular, near--solar metallicity
galaxy, it is possible that high wind mass--loss rates are encountered for massive rapidly rotating stars in this environment.
Another open question with regards to models CSM0\_A and CSM2\_A remains 
the UV--bright nature of the plateau luminosity and whether radiation from the spin--down of a young magnetar can account for it.

\subsection{{\it Pulsational PISN and Hydrogen--Poor CSM Interaction.}}\label{csmonly}

Another possiblity for ASASSN--15lh is a pure H--poor CSM interaction scenario involving the collision
of massive shells ejected by a PPISN \citep{2007Natur.450..390W,2012ApJ...760..154C,2016arXiv160204865S}.
A problem with a pure PPISN shell collision scenario is the implied energetics; simulations of PPISNe
yield kinetic energies of the order of $10^{51}$~erg for the shells over time--scales of months to a year implying
luminosities up to $\sim 10^{44}$~erg~s$^{-1}$ \citep{2007Natur.450..390W,2016arXiv160204865S}. In light of these results, the
only possibility would be interaction of massive SN ejecta by an energetic CCSN following the last 
PPISN shell ejection. The explosion would then form a few $M_{\odot}$ of $^{56}$Ni and could leave a black hole behind 
but the bulk of the luminosity would be provided by the forward and the reverse shocks following CSM interaction.

Models CSM0 and CSM2 in the lower two panels of Figure~\ref{Fig:LCfits} show pure CSM interaction 
fits to the LC of ASASSN--15lh. As was the case
for model CSM2\_B, model CSM2 fails to reproduce the late--time plateau leaving CSM0 as the only
viable possibility thus suggesting interaction with a dense shell.
For this model, 36~$M_{\odot}$ of H--poor SN ejecta interacts with an also H--poor 19.5~$M_{\odot}$ CSM
for a total mass of $\sim$~55.5~$M_{\odot}$ consistent with a rapidly--rotating PPISN progenitor. The
derived $\dot{M}$ of 0.2~$M_{\odot}$~yr$^{-1}$ suggest that the shell ejection occured $\sim$~100~years
prior to the SN. This timescale is consistent with some of the timescales
between PPISN pulses listed in Supplementary Table 1 of \citet{2007Natur.450..390W} for non--rotating 
progenitors in the mass range 54--56~$M_{\odot}$ in agreement with the parameters implied by the CSM0 model LC fit. 
The derived radius of the CSM shell ($R_{\rm CSM} = 5 \times 10^{15}$~cm) is also consistent with the
black body fits presented by \citet{2016Sci...351..257D}.


One common objection to models of H--poor CSM interaction for SLSN--I is the absence of interaction
features like those seen for SLSN--II (luminous SN IIn) events. We stress, however, that more 
numerical work and updated algorithms
need to be implemented in non--LTE radiation transfer codes to carefully evaluate the conditions that allow
emission line formation for elements other than H and He for non--homologous, non--monotonic 
velocity profiles \citep{2013ApJ...773...76C,2015MNRAS.449.4304D,2016MNRAS.tmp..117D}. 

\section{DISCUSSION AND CONCLUSIONS}\label{Disc}

In this paper we studied physically plausible SN models that may account for the LC of the record--breaking SLSN--I 
ASASSN--15lh (or SN~2015L; \citealt{2016Sci...351..257D}). Our goal was to fit the complete LC of
the event, including the latest ($t >$~100~d) observations that show a UV--bright plateau phase
suggesting continuous heating of the SN ejecta. 

To that goal we employed semi--analytic ``hybrid'' models that consider the contribution of
all three power--input mechanisms discussed for SLSNe: the radioactive decays of $^{56}$Ni and 
$^{56}$Co, energy injection by the spin--down of a newly born magnetar and CSM interaction.
We found that in all plausible cases the contribution of $^{56}$Ni input is negligible, therefore
we focused on models of magnetar spin--down and forward and reverse shock heating
following H--poor CSM interaction. 
Also, we studied cases where the late--time plateau is powered either by the magnetar input
or by reverse shock heating and discussed implications with regards to the observed UV--bright
flux during this phase.

We found that models that involve interaction with a massive steady--state wind (CSM2\_B and CSM2)
fail to reproduce the late plateau phase with the exception of model CSM2\_A where the magnetar
input supplies the luminosity deficit for $t >$~80~d. 

\begin{figure}       
\hskip -0.03 in
\centerline{\psfig{figure=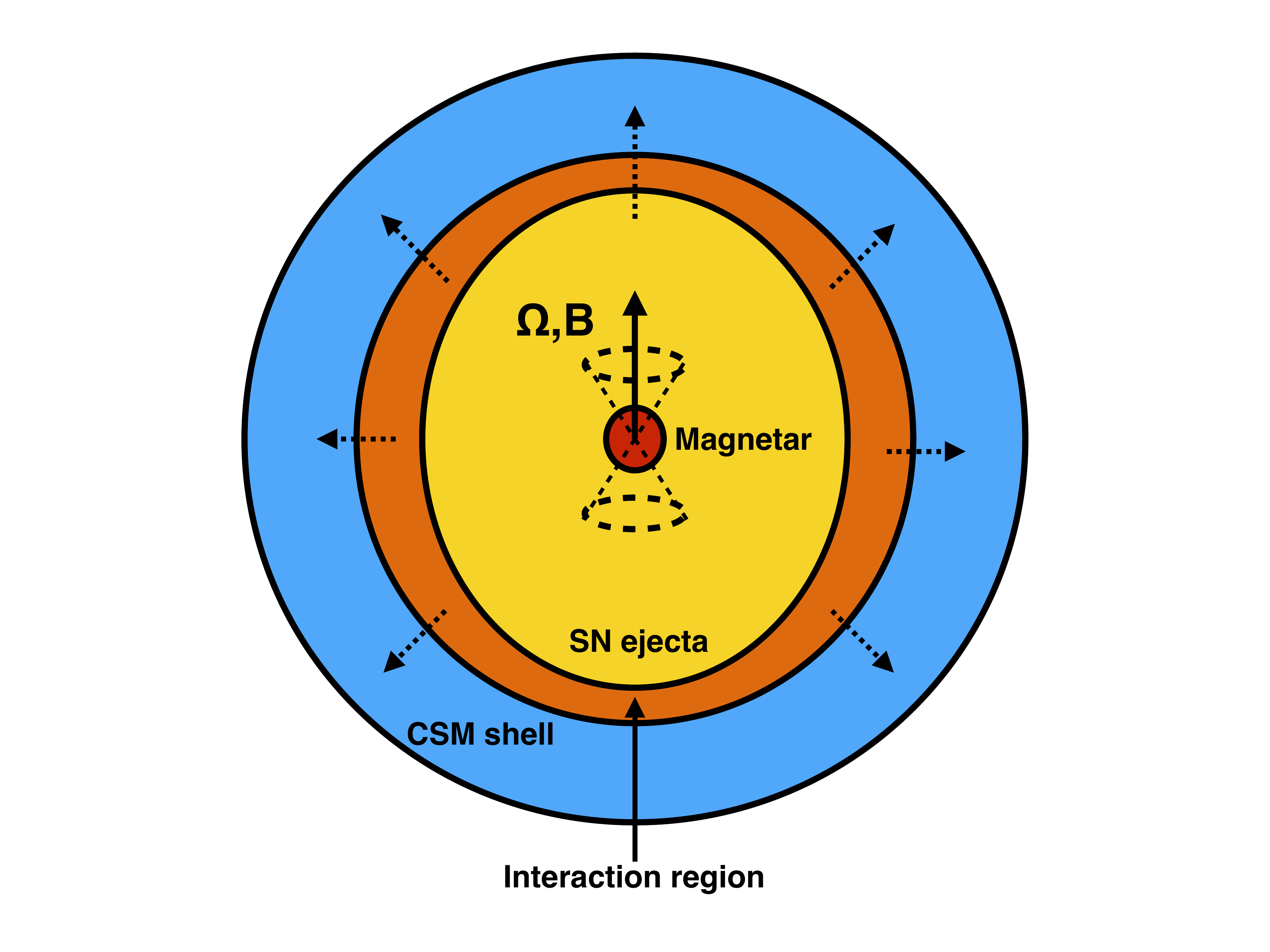,angle=0,width=10cm,height=7cm,trim=0.3in 0.6in 0.5in 0.25in,clip}} \\
\hskip 1.0 in
\centerline{\psfig{figure=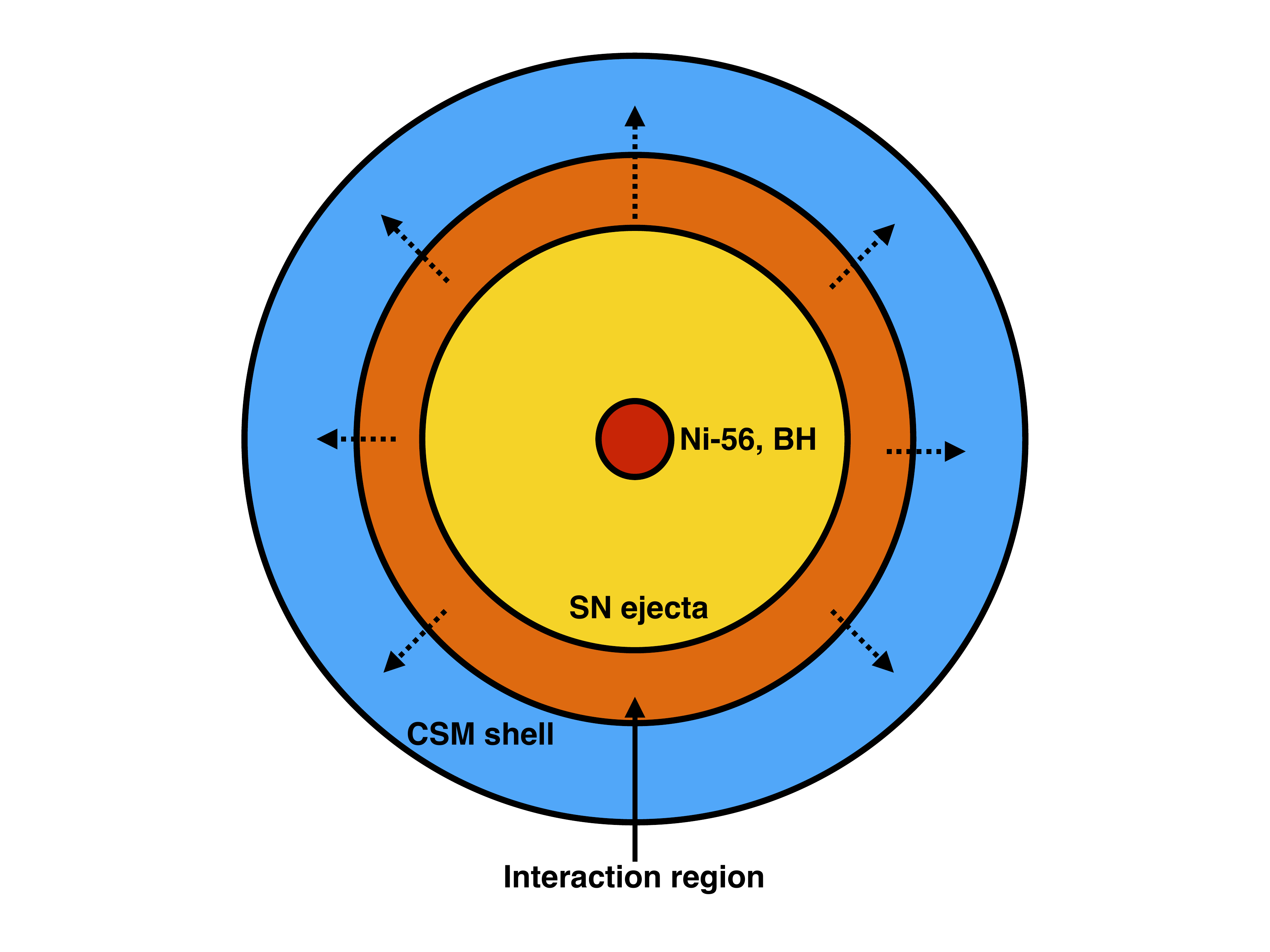,angle=0,width=10cm,height=7cm,trim=0.3in 0.6in 0.5in 0.25in,clip}} \\
\hskip 1.0 in
\centerline{\psfig{figure=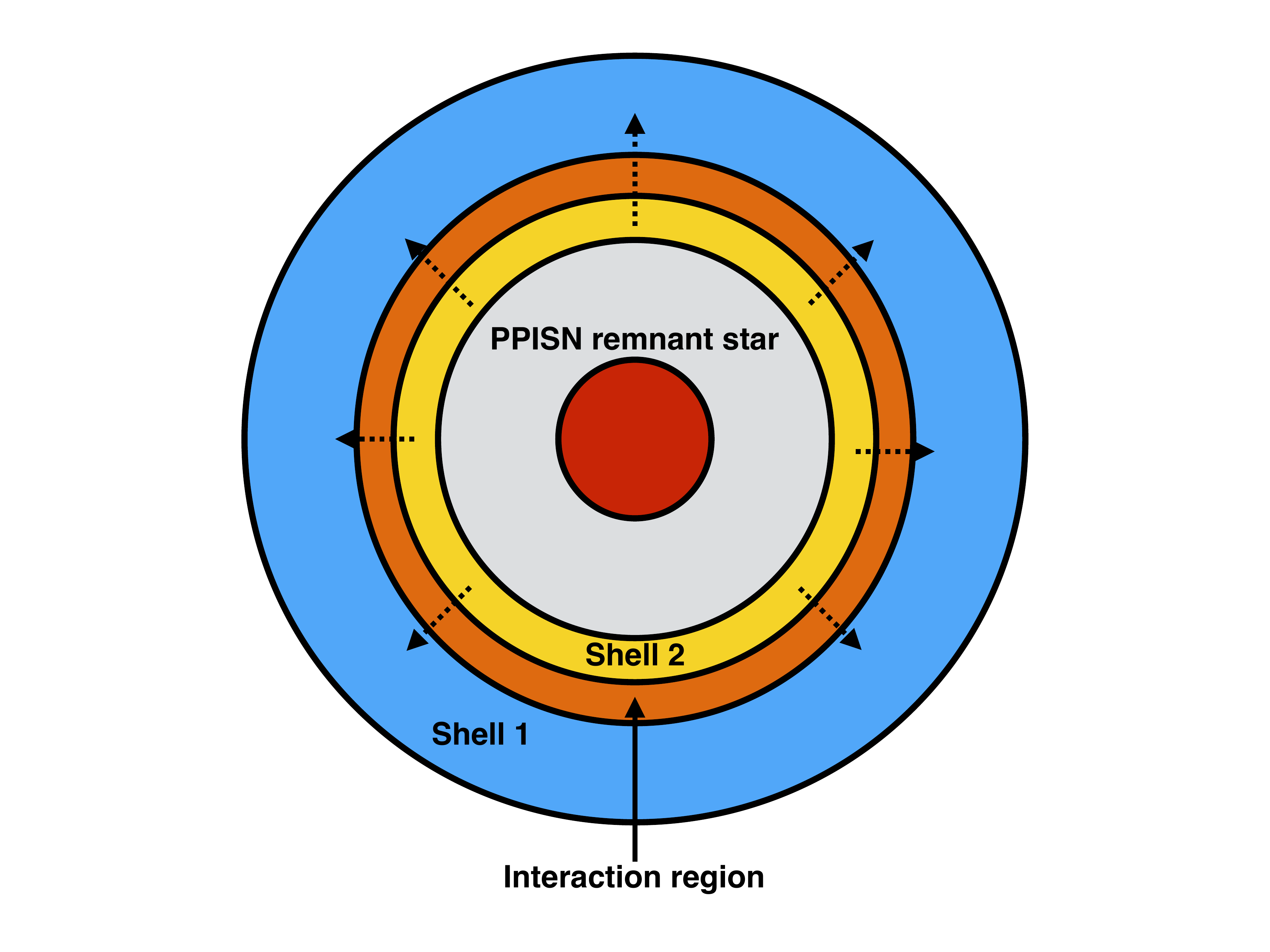,angle=0,width=10cm,height=7cm,trim=0.3in 0.6in 0.5in 0.25in,clip}}
\caption{Schematic diagrams of three possible progenitor models for ASASSN--15lh. {\it Upper panel:} Massive, rapidly rotating CCSN
explosion following shell ejection due to a PPISN. Radiative energy input is supplied by the spin--down of a newly--born magnetar in the center
combined with H--poor CSM interaction of the SN ejecta with the PPISN shell. Rapid rotation is required for magnetar formation that can lead
to asymmetric SN ejecta (\S~\ref{models}).
{\it Middle panel:} Interaction of H--poor SN ejecta from a massive CCSN following shell ejection due to a PPISN. The core of the progenitor
star in this case collapses to a black--hole and produces $\sim$~1-4~$M_{\odot}$ of $^{56}$Ni.
{\it Lower panel:} H--poor interaction between massive PPISN shells perhaps followed by an energetic SN explosion.
\label{Fig:cartoon_models}}
\end{figure}

On the contrary, models that invoke interaction with a massive dense shell lost via an eruptive
mass-loss mechanism (CSM0\_A, CSM0\_B and CSM0) 
provide the best fits to the LC of ASASSN--15lh.
Nevertheless, there are issues with the hybrid models invoking a magnetar input since it requires the
presence of a magnetar and more fine--tuning to fit the LC of ASASSN--15lh. There are also yet unresolved issues with efficient thermalization
of magnetar radiation in the SN ejecta and conversion to luminosity discussed in \S~\ref{ppisnmag}. 
As such, the more consistent and better constrained single--input model CSM0 where the reverse shock provides the luminosity 
for the late--time plateau is favored in our analysis.
The derived parameters from our fits suggest a common theme: interaction of massive SN ejecta
($M_{\rm SN} \simeq$~36~$M_{\odot}$ with a H--poor CSM shell of $\sim$~20~$M_{\odot}$). The derived
total SN ejecta and CSM masses of our models support a scenario where a rapidly--rotating 
energetic CCSNe exploded within a previously shed massive H--deficient shell ejected either via
the PPISN \citep{2007Natur.450..390W} or the LBV \citep{2006ApJ...645L..45S} mechanism. 
Schematical diagrams of the proposed alternative progenitor and power--input configurations
for ASASSN--15lh are shown in Figure~\ref{Fig:cartoon_models}.

A rapidly rotating ($>$~50\% of the break--up speed at the equator)
$\sim$~50-60~$M_{\odot}$ star may undergo enhanced mixing allowing
it to reach the PPISN regime during its core oxygen burning phase \citep{2013ApJ...776..129C}.
At this phase, the star may have lost the entirety of its H and He envelope due to enhanced mass--loss
because of rotation, duplicity or strong winds as expected for the host of ASASSN--15lh that is a
bright, near--solar metallicity galaxy. 
Upon undergoing the PPISN, the bare CO core of the star can eject several solar masses of H--poor
material forming a massive shell around the remnant. The core remnant can then evolve to become
an energetic ($\sim 10^{52}$~erg) CCSN explosion inevitably leading to strong CSM interaction. 

The forward and reverse shock heat deposition in the H--poor CSM shell and SN ejecta could
be supplemented by radiation from a newly--born rapidly--rotating magnetar
($B_{\rm mag} = (1.1-1.2) \times 10^{13}$~G and $P_{\rm mag} =$~1~ms). Upon the collapse
of such massive progenitor, black hole formation may be more likely, but alternative
channels exist to allow for magnetar birth \citep{2006ApJ...636L..41M,2012ApJ...757...69U,2014ApJ...783...10S,2016arXiv160300511W}. 
The fact that the plateau phase is UV--bright, however,
favors a shock heating input for the late luminosity of ASASSN--15lh consistent with a single--input strong CSM interaction scheme.

Current and next generation transient searches like the Zwicky Transient Factory ({\it ZTF}), {\it PanSTARRS}
and {\it LSST} will likely yield more extraodinary events like ASASSN--15lh that
put the known SN power engine models to the test. In addition, 
radio observations of SLSN--I may help distinguish between the different power--input mechanisms \citep{2016arXiv160304748N}.
In tandem, advances in numerical algorithms and parallel
computing will eventually allow us to accurately model the spectra of SLSN--I and be a step
closer to understanding the extreme origins of extreme supernovae.

\acknowledgments

The research of EC and JCW is supported by NASA grant HST-AR-13276.002-A.
EC thanks the Enrico Fermi Institute for its support via the Enrico
Fermi Fellowship.


\bibliography{massiverot}

\end{document}